\documentclass[12pt]{iopart}
\usepackage{iopams}
\usepackage{graphicx}
\usepackage{bm}

\newcommand{\Z}{\mathbb{Z}}
\newcommand{\N}{\mathbb{N}}
\newcommand{\beq}{\begin{equation}}
\newcommand{\eeq}{\end{equation}}

\newcommand{\n}{\mathbf{n}}

\newcommand{\x}{\mathbf{x}}
\newcommand{\eps}{\varepsilon}
\newcommand{\mafrac}{\sqrt{\frac{8 \pi}{3}}}

\begin{document}
\title[Observables in a lattice Universe]{Observables in a lattice Universe.\\
{\it The cosmological fitting problem.}}
\author{Jean-Philippe Bruneton$^1$ and Julien Larena$^2$}
\address{$^1$ Namur Center for Complex systems (naXys), University of Namur, Belgium}
\address{$^2$ Department of Mathematics, Rhodes University, Grahamstown 6140, South Africa.}
\ead{jpbr@math.fundp.ac.be; j.larena@ru.ac.za}
\begin{abstract}
We explore observables in a lattice Universe described by a recently found solution to Einstein field equations. This solution models a regular lattice of evenly distributed objects of equal masses.
This inhomogeneous solution is perturbative, and, up to second order in a small parameter, it expands at a rate exactly equal to the one expected in a dust dominated Friedmann-Lema\^itre-Robertson-Walker (FLRW) model with the equivalent, smoothed, energy density. Therefore, the kinematics of both cosmologies are identical up to the order of perturbation studied.
Looking at the behaviour of the redshift and angular distance, we find a condition on the compactness of the objects at the centre of each cell under which corrections to the FLRW observables remain small, i.e. of order of a few percents at most. Nevertheless, we show that, if this condition is violated, i.e. if the objects are too compact, our perturbative scheme breaks down as far as the calculations of observables are concerned, even though the kinematics of the lattice remains identical to its FLRW counter-part (at the perturbative order considered). This may be an indication of an actual fitting problem, i.e. a situation in which the FLRW model obtained from lightcone observables does not correspond to the FLRW model obtained by smoothing the spatial distribution of matter. Fully non-perturbative treatments of the observables will be necessary to answer that question.
\end{abstract}
\section{Introduction}
It has long been recognised that calculating observables in an inhomogeneous Universe could be quite challenging. Indeed, our usual description of the geometry of the Universe on large scales relies on Friedmann-Lema\^itre-Robertson-Walker (FLRW) models in which the distribution of matter is assumed perfectly homogeneous and isotropic. As a result, light rays propagate in a homogeneous medium filled with matter and are therefore sensitive only to the Ricci curvature of spacetime. In the real Universe, on the other hand, at least in the late stages of its evolution, matter is clumped into virialised objects with large (almost) empty regions between them, and light therefore travels mainly in empty space, where its behaviour is dominated by the Weyl curvature of spacetime, rather than the Ricci curvature. This raises the natural question to whether the FLRW approximation is suitable to calculate observables in the late time Universe: under which conditions is it possible to replace Weyl curvature along the line of sight with an 'equivalent' Ricci curvature? And what is this 'equivalence' all about? The problem is therefore what has been dubbed a 'fitting problem' \cite{Ell84, Ellis:1987zz}, that is: how do we replace the real Universe by an FLRW idealisation? In principle, this leads to a backreaction issue: the idealised FLRW given by observations will not, in general be the FLRW model for which the actual matter density has simply been smoothed in space. In the standard model, this is usually accounted for via the Dyer-Roeder equation \cite{Dyer:1973zz} (see \cite{1969ApJ...155...89K} for earlier results on this problem), but recent works tend to show that there might be better ways to model the effect \cite{Clarkson:2011br, Bolejko:2010nh,Kainulainen:2009dw,Kainulainen:2010at}. This issue has been addressed by a certain number of authors lately; see e.g., \cite{Bolejko:2010nh, Meures:2011gp,Clarkson:2011br,Bolejko:2012ue}. Usually, they use a 'realistic' model of structure formation, either through cosmological perturbation theory or N-body simulation, to evaluate the impact of inhomogeneities on the propagation of light; others have used Swiss-Cheese models \cite{Brouzakis:2007zi,Marra:2007pm,Vanderveld:2008vi,Flanagan:2011tr}, mainly to see whether inhomogeneities along the line of sight could account for the Dark Energy phenomenon. Our take on the problem is slightly different. In particular, we are not trying to address the Dark Energy problem; we simply try to understand under which conditions, in the controlled environment of an (almost) exact solution to Einstein field equations, the behaviour of null geodesics can safely be approximated by the null geodesics of an FLRW model.\\
\\
In a previous paper \cite{Bruneton:2012cg}, we proposed a lattice solution to Einstein field equations made of equal masses $M$ separated by a comoving distance $L$. This solution is accurate at order $M/L$, and can be expanding or contracting. We proved that this solution is kinematically equivalent (at the order of perturbation considered) to an FLRW model with a dust matter content having an energy density equal to the one obtained by smoothing the lattice distribution, i.e. $M/L^{3}$. This was interpreted as supporting the usual fluid approximation in cosmology. Recently cosmological lattices similar to the one presented in \cite{Bruneton:2012cg} have been studied both analytically \cite{Clifton:2012qh} and numerically \cite{Yoo:2012jz}. The lattice studied in \cite{Yoo:2012jz} is similar to the one we studied, with cubic symmetry, and their results agree with ours, since they recover an Einstein-de Sitter Universe in the limit $M/L\ll 1$ (large separation between the masses). Similarly, the results of \cite{Clifton:2012qh} show that, for an infinite number of masses, the kinematics of the lattice tends towards the kinematics of the Einstein-de Sitter Universe.\\ 
The solution in \cite{Bruneton:2012cg}, being kinematically equivalent to an FLRW model at the perturbative order considered, is the ideal setting to study observables: if observables only deviate slightly from the observables in the kinematically equivalent FLRW Universe, then we might say that this FLRW model is a good fitting model. On the other hand, if observables exhibit large variations compared to the one in the analogue FLRW, that will show that a perturbative calculation is unable to describe accurately observables in the lattice, even though it is a good approximation to the geometry of spacetime; in that case, further, non perturbative studies will be needed.\\
\\
The propagation of light in lattice models has been studied before \cite{Clifton:2009jw,Clifton:2011mt} in the context of Lindquist-Wheeler models \cite{LindWheel}. These models are only approximate, and the propagation of light through the boundaries between cells is not fully controlled, leading to important differences in the results, depending on the approximation scheme used \cite{Clifton:2011mt}. Our model, on the contrary, is only approximate in power of $M/L$, and not on the way cells are glued together (much like the model developed in \cite{Clifton:2010fr}). Therefore, propagating light through the lattice is not a problem. The only limitation will come from the fact that we can only trust the solution for a small range of redshifts (typically $z\ll 1$). By solving Sachs equations at order $M/L$, we will see that at this order the equations for the shear and the isotropic expansion decouple. This implies that at order $M/L$, {\it a priori}, the Weyl curvature of spacetime does not play any role in the calculation of the distance/redshift relation. Then, the formal solution appears to be equivalent to the solution in the analogue FLRW model, plus small corrections of order $M/L$. Nevertheless, by studying carefully the order of magnitude of these corrections, we will prove that they remain small only under specific conditions on the compactness of the objects forming the lattice: when these objects are too compact, the perturbative expansion breaks down, and the differences between observables in the lattice and in the analogue FLRW model cannot be simply evaluated by using perturbative methods. That might indicate a tension between the fitting models constructed kinematically and observationally.
This paper somehow illustrates, on a specific example, the concepts defined in \cite{Kolb:2009rp}: using the terminology and classification introduced in that paper, in our lattice Universe, the strong backreaction is absent and the Global and Averaged Background Solutions (GBS and ABS) coincide at the perturbative order considered here, but weak backreaction are present and the Phenomenological Background Solution (PBS) does not necessarily coincide with the others. Indeed, on the one hand, if our compactness criterion is satisfied, the difference between the PBS and the GBS (and ABS) is small, of order of a few percents at most. On the other hand, when objects in the lattice are too compact, perturbative calculations cannot be trusted to give the structure of the PBS, and fully non-perturbative calculations must be done to evaluate by how much this PBS differs from the ABS (and from the GBS). The break-down in the perturbative calculations can be attributed to the fact that the shear of the ray bundles cannot be neglected anymore.\\
\\
The paper is organised as follows. Section~2 will briefly present the solution described in details in \cite{Bruneton:2012cg}. Section~3 details the calculations of the redshift and angular distance in the lattice Universe. For the sake of clarity, Section~3 focusses only on the analytical expressions, while their physical interpretation, numerical calculations, and discussion of the effect of inhomogeneities on the propagation of light, especially in regards of the fitting problem and of the Weyl focussing, are left for Section~4. Finally, section~5 will be a conclusion. \\
\\
Throughout the paper, the signature of the spacetime metric will be $(-,+,+,+)$, and, unless otherwise specified, we will work in units of $c=G=1$.

\section{The lattice solution}
In a previous paper \cite{Bruneton:2012cg} we found that an infinite, regular, cubic lattice of masses $M$ spaced by a comoving length $L$, in an otherwise empty Universe without any cosmological constant, is described by the following metric in a synchronous comoving coordinate system:
\begin{eqnarray}
\label{mainresultmetric}
g_{00}&=&-1, \, g_{0i}=0, \nonumber \\
g_{ij}&=& \delta_{ij} \left[1+ 2\eps\sqrt{\frac{GM}{Lc^2}} \mafrac \frac{c t}{L} +\frac{2GM}{L c^2}\left(f(\mathbf{x})+ \frac{2 \pi c^2 t^2}{3 L^2} \right) \right] \nonumber\\
&+&\frac{G M}{L c^2} c^2 t^2\partial_{ij}^2 f(\mathbf{x})+ \mathcal{O}\left(\frac{M^{3/2}}{c^{3}L^{3/2}}\right),
\end{eqnarray}
up to order $GM/Lc^2$, and where $f$ is given in Eq.~(\ref{fdef2}) below. This perturbative, approximate solution, is justified phenomenologically by the fact that a lattice of galaxies separated by the typical intergalactic distance today have $G M/L c^2 \sim 10^{-8}$. In the above metric, $\eps =\pm 1$ corresponds respectively to an expanding or contracting lattice. In the following we shall restrict ourselves to the expanding lattice only ($\eps=1$). This metric solves Einstein equations without cosmological constant up to order $M/L$, with the source term associated to the lattice, namely:
\beq
\label{source1}
T_{00} = \frac{M}{L^3}\sum_{\n \in \Z^3} e^{i \frac{2 \pi \mathbf{n}.\x}{L}} +\mathcal{O}\left(\frac{M^{3/2}}{L^{3/2}}\right), \quad T_{0i} = T_{ij}= \mathcal{O}\left(\frac{M^{3/2}}{L^{3/2}}\right), 
\eeq
where the point-masses are described by a three-dimensional Dirac comb. In this case, the function $f$, accounting for the anisotropies of the gravitational field created by the masses, reads:
\beq
\label{fdef2}
f(\x)=\frac{1}{\pi}\sum_{\mathbf{n}\in\Z^{3}_{*}}\frac{e^{\frac{2\pi}{L} i\mathbf{n}.\x}}{|\mathbf{n}|^{2}}=\frac{8}{\pi}\sum_{(n,p,q)\in\N^{3}_{*}}\frac{\cos\left(\frac{2\pi}{L}nx\right)\cos\left(\frac{2\pi}{L}py\right)\cos\left(\frac{2\pi}{L}qz\right)}{n^{2}+p^{2}+q^{2}},
\eeq
where $\N^{3}_{*}$ stands for $\N^{3}\setminus \left\{\left(0,0,0\right)\right\}$ and similarly for $\Z^3_{*}$. This solution presents singularities at the position of the masses as a natural consequence of modelling them with a Dirac distribution. However the solution can be regularized in the UV, e.g. by describing instead the lattice of masses by a three-dimensional lattice of peaked Gaussians of width $\eta$. In this case the non-vanishing part of the stress energy tensor now reads
\begin{equation}
T_{00}=S(x)S(y)S(z),
\end{equation}
with:
\beq
S(x)=\sum_{n \in \Z}\frac{1}{\eta \sqrt{\pi}}e^{-\frac{(x-n L)^{2}}{\eta^{2}}},
\eeq
and similarly for $S(y)$ and $S(z)$. This source term can be written in Fourier space as:
\beq
\label{source2}
T_{00}=\frac{M}{L^{3}}\sum_{\mathbf{n}\in\Z^3}e^{\frac{2\pi}{L}i\mathbf{n}.\x-\frac{\pi^{2}|\mathbf{n}|^{2}\eta^{2}}{L^{2}}}+ \mathcal{O}\left(\frac{M^{3/2}}{L^{3/2}}\right).
\eeq
Then, the above metric, Eq.~(\ref{mainresultmetric}), is still a solution to Einstein equations provided the function $f$ is now replaced by the following regularized function $f_{\eta}$:
\begin{eqnarray}
f_{\eta}(\x)&=&\frac{1}{\pi}\sum_{\mathbf{n}\in\Z^{3}_{*}}\frac{e^{-\frac{\pi^{2}|\mathbf{n}|^{2}\eta^{2}}{L^{2}}}}{|\mathbf{n}|^{2}}e^{\frac{2\pi}{L}i\mathbf{n}.\x} \nonumber \\
\label{feta}&=&\frac{8}{\pi}\sum_{(n,p,q)\in\N^{3}_{*}}\frac{e^{-\frac{\pi^{2} (n^2+p^2+q^2)\eta^{2}}{L^{2}}}\cos\left(\frac{2\pi}{L}nx\right)\cos\left(\frac{2\pi}{L}py\right)\cos\left(\frac{2\pi}{L}qz\right)}{n^{2}+p^{2}+q^{2}}.
\end{eqnarray}
An important feature of both the source terms presented in Eqs.~(\ref{source1}) and (\ref{source2}) is the fact that the infinite wavelength mode (the zero mode $\mathbf{n}=0$ of the Fourier expansion of the source), behaves as an homogeneous comoving density $M/L^3$. This Friedmannian-like component for the source precisely produces  in Eq.~(\ref{mainresultmetric}) the FLRW metric of a flat, dust-dominated Universe, when expanded in powers of $H_0 = \sqrt{8 \pi M/3L^{3}}$, and up to order $H_0^2$, see \cite{Bruneton:2012cg}. On the other hand, finite wavelength modes of the source distribution have no Friedmannian counterparts and account for its inhomogeneity. These modes generate the terms proportional to $f_{\eta}$ and its derivatives in Eq.~(\ref{mainresultmetric}). The metric of the lattice Universe is thus given by the FLRW metric, truncated at the correct order, plus corrections coming from the inhomogeneity and anisotropy of the source distribution. As a consequence, we expect that the observables in the lattice Universe will partially match the ones in FLRW, up to the degree of approximation considered here, but with corrections coming from the anisotropic terms in the metric. Therefore, our main task in this paper will be to assess the amplitude of these corrections with respect to the FLRW behaviour at the level of observable quantities, such as redshift and angular distance.\\
\\
The validity of the approximations made depend essentially on two dimensionless parameters, $M/L$ and $\eta/L$; their ratio is simply the compactness of the objects. As mentioned above, the numerical values of $M$ and $L$ can be chosen to be of the order of magnitude of the mass of a galaxy and of a typical intergalactic distance:
$$
M\sim 10^{11} M_{\odot} \mbox{ and } L\sim 1 \mbox{ Mpc},
$$
where $M_{\odot}$ is a solar mass. The choice of $\eta$, the spread of the objects, is more difficult to make. A natural cut-off is provided by the Schwarzschild radius of the masses: $\eta \sim 2 M$. However, in the cosmological context, it is more relevant, as well as much easier for numerical purposes, to choose a factor $\eta$ that matches with the size of the objects considered here, i.e. the size of a typical galaxy, which is much larger than its Schwarzschild radius. In that case, a good choice would be: $\eta \sim 10$ kpc, so that $\eta/L \sim 0.01$. With this value of $\eta$, the UV-regulator $e^{-\frac{\pi^{2} (n^2+p^2+q^2)\eta^{2}}{L^{2}}}$ makes the various sums converge quite quickly, and we shall thus restrict ourselves to a large but finite number of terms (typically of order $200^{3}$) in the expression of the function $f_{\eta}$, when we will deal with numerical studies in the next section.\\
\\
Note that this choice of numerical values is only indicative. In the rest of the paper, we will keep $M/L$ fixed, but $\eta/L$ will be allowed to change, since the main result of the paper is the fact that observables, when calculated perturbatively, only remain close to their FLRW analogues provided:
\beq
\label{mainboundIntro}
\frac{M}{L} \ll \mathcal{O}(1) \times \left(\frac{\eta}{L}\right)^4,
\eeq
i.e. provided the objects are not too compact. If this bound is not satisfied, the perturbative expansion used in this paper fails to be well-defined, and other, non-perturbative techniques should be used to address the fitting problem. Note that the choice of $\eta$ mentioned earlier, $\eta/L\sim 0.01$, that corresponds to the value expected for a galaxy-like object, is marginally violating condition~(\ref{mainboundIntro}), since it corresponds to the case $M/L\sim\left(\eta/L\right)^{4}$.\\
\\
In the following we shall also use some shortcuts for the metric, which we write formally as a power series $g_{ab}=\eta_{ab}+\sqrt{\frac{M}{L}}\delta g_{ab}+ \frac{M}{L}\delta^2 g_{ab}+ \mathcal{O}\left(\frac{M^{3/2}}{L^{3/2}}\right)$, where $\eta_{ab}$ is the Minkowski metric, and $\delta g_{ab}$ and $\delta^2 g_{ab}$ correspond respectively to the order $\sqrt{M/L}$ and $M/L$ parts of the metric in Eq.~(\ref{mainresultmetric}). Accordingly, the Christoffel symbols are decomposed into $\Gamma^{a}_{bc}=\sqrt{\frac{M}{L}}\delta \Gamma^{a}_{bc}+ \frac{M}{L}\delta^2 \Gamma^{a}_{bc}+ \mathcal{O}\left(\frac{M^{3/2}}{L^{3/2}}\right)$.

\section{Observables: analytical expressions for redshift and distance}
In order to evaluate observables, we investigate the properties of null geodesics in the lattice Universe. In the following, $k^a$ refers to the dimensionless four momentum of a null ray $k^a=d x^a/ d\lambda$, where $\lambda$ is an affine parameter along the ray and has the dimension of a length. In Section~3.1 we solve perturbatively the geodesic equations in order to find the redshift $z(\lambda)$ as a function of the arrival direction of the photon at the observer, see Eq.~(\ref{redshift}) below. In Section~3.2, we then use and solve perturbatively the Sachs optical equations for a bundle of light rays in the lattice Universe. This enables us to compute the expansion $\theta$ of the light-bundle, and thus the angular distance as a function of the affine parameter $r_A(\lambda)$, see Eq.~(\ref{ral}) below. In this section, we will provide analytic expressions for $z(\lambda)$ and $r_{A}(\lambda)$. The discussion of the amplitude of each term in these expressions is left for Section~4. We shall work first in natural units and re-establish dimensionful constants $c$ and $G$ at the end of the calculation. 

\subsection{Geodesic equation for light rays}
The geodesic equation reads:
\beq
\frac{D k^a}{D \lambda} = \frac{d k^a}{d \lambda} + \Gamma^{a}_{bc} k^b k^c = 0,
\eeq
together with $k_a k^a=0$. We look for a pertubative solution of the form:
\beq
k^a(\lambda)=v^a(\lambda)+\sqrt{\frac{M}{L}} \xi^a(\lambda)+\frac{M}{L} \zeta^a(\lambda)+\mathcal{O}\left(\frac{M^{3/2}}{L^{3/2}}\right).
\eeq
The zeroth order is simply $\dot{v^a}=0$, where a dot refers to $d/d\lambda$. We pick up the solution $v^a=(1, v_x, v_y, v_z)$ with $v_x^2+v_y^2+v_z^2=1$, so that the $v^i$'s shall represent the arrival direction of the photon at the observer (see below). The first order equations (in $\sqrt{M/L}$) then read: $\dot{\xi^a}+ \delta \Gamma^{a}_{bc}v^b v^c=0$. From the metric Eq.~(\ref{mainresultmetric}), this explicitly reads: 
\begin{eqnarray}
\dot{\xi^0}= -\mafrac \frac{1}{L}, \, \, \textrm{and} \, \, \,\dot{\xi^i}= -2 \mafrac \frac{v^i}{L}.
\end{eqnarray}
These equations are solve by:
\begin{eqnarray}
\xi^0(\lambda)&=&-\frac{\lambda}{L}\mafrac \\
\xi^i(\lambda)&=&-\frac{2 v^i \lambda}{L}\mafrac,
\end{eqnarray} 
provided the boundary conditions are set such that the observer is located at $\lambda_0=0$ and $t_0=0$. Then $\lambda$ is increasing with the cosmic time and is negative along the past light cone of the observer. We will take similar conventions for $\zeta^{a}$, so that we will have $k^a(\lambda_0)=(1, v_x, v_y, v_z)$, and the $v_i$'s thus give the direction of the light ray arriving at the observer, as announced. This solution can then be used to deduce the expression for the look-back time to the leading order:
\beq
-t= -\lambda + \sqrt{\frac{M}{L}}\mafrac \frac{\lambda^2}{2L} +\mathcal{O}(M/L).
\eeq
Notice that the expression for the look-back time does not involve the usual present time $t_0$. Actually the fact that $t_0=0$ comes from the constraint $k_a k^a=0$ at first order, as one might check directly. Accordingly, the present time $t_0$ was set to zero by convention in the metric Eq.~(\ref{mainresultmetric}).  The reason behind the convenience of this choice is that there is no notion of absolute time in the lattice Universe, as the dynamical evolution of the lattice solution into the past cannot be followed until it reaches any singularity. Indeed, denoting $L_{phys}(t)$ the physical distance between the masses (basically equal to $L$ times an effective scale factor, see \cite{Bruneton:2012cg}), the approximation $M/L_{phys}(t) \ll 1$ would break down at a certain point in the past, because $L_{phys}$ gets smaller and smaller.\\
\\
Assuming that the observer is located at $x^i=0$, we also have:
\beq 
x^i(\lambda) = v^i\lambda +\mathcal{O}(\sqrt{M/L}).
\eeq
The second order equation then reads:
\beq
\dot{\zeta}^a+ 2 \delta \Gamma^{a}_{bc} v^b \xi^c+ \delta^2 \Gamma^{a}_{bc} v^b v^c=0,
\eeq
where it is enough, at this order, to replace $t$ by $\lambda$ and $x^i$ by $v^i \lambda$. Then, the temporal component of the geodesic equation reads:
\beq
\dot{\zeta}^{0}(\lambda)+\lambda \partial_{ij} f_{\eta}(\x(\lambda)) v^i v^j = \frac{28 \lambda \pi}{3 L^2},
\eeq
where summation is meant on repeated Latin indices, and $f_{\eta}$ was given by Eq.~(\ref{feta}). One may derive a similar equation for $\dot{\zeta}^i$ and also check that the constraint $k_a k^a=0$ is satisfied to this order. As far as the redshift is concerned however, we only need $k^0$ and thus $\zeta^0$.  Noticing then that: 
\beq
-\frac{d}{d \lambda} \left[f_{\eta}(\x(\lambda))-\lambda \partial_{i} f_{\eta}(\x(\lambda)) v^i\right]=\lambda \partial_{ij} f_{\eta}(\x(\lambda)) v^i v^j,
\eeq
the differential equation for $\zeta^0$ is solved by:
\beq
\zeta^0(\lambda_0)-\zeta^0(\lambda)=\left[f_{\eta}(\x(\lambda))-\lambda \partial_{i} f_{\eta}(\x(\lambda)) v^i\right]^{\lambda_0}_{\lambda} + \left[\frac{14 \pi \lambda^2}{3 L^2} \right]^{\lambda_0}_{\lambda}.
\eeq
With the boundary conditions explained previously, namely $\lambda_0=0$ and $\zeta^0 (0)=0$, we finally get:
\beq
\zeta^0(\lambda)= \frac{14 \pi \lambda^2 }{3 L^2}+\left[ f_{\eta}(\x(\lambda))-\lambda \partial_{i} f_{\eta}(\x(\lambda)) v^i\right]^{\lambda}_{0},
\eeq
The redshift can now be deduced using:
\begin{equation}
1+z(\lambda)=\frac{\left(k^{a}u_{a}\right)_{S}}{\left(k^{a}u_{a}\right)_O}=\frac{k^0 (\lambda)}{k^0(\lambda_0)}= k^0(\lambda)=1+\sqrt{\frac{M}{L}} \xi^0(\lambda)+\frac{M}{L} \zeta^0(\lambda),
\end{equation}
up to order $M/L$, where the subscripts $S$ and $O$ refer respectively to the source and the observer, and where we used that fundamental observers, such as the masses themselves in our case, have $u^a=(1,0,0,0)$. We also used that $k^0(\lambda_0)=1$ as a consequence of the normalisation chosen. We thus get the following law for the redshift:
\begin{eqnarray}
\label{redshift}
z(\lambda)&=& -\sqrt{\frac{GM}{Lc^2}}\mafrac \frac{\lambda}{L} \nonumber \\ &+&\frac{GM}{Lc^2}\left(\frac{14 \pi \lambda^2 }{3 L^2}+\left[ f_{\eta}(\x(\lambda))-\lambda \partial_{i} f_{\eta}(\x(\lambda)) v^i\right]^{\lambda}_{0}\right) +\mathcal{O}\left(\frac{M^{3/2}}{L^{3/2}}\right).
\end{eqnarray}
This expression coincides with its FLRW counterpart for a flat, dust-filled Universe without any cosmological constant, up to order $H_0^2 \propto G M/L^3 c^2$. The anisotropies only show up in the term into brackets, and at second order in $\sqrt{M/L}$. This formula is further discussed in Section 4.1.

\subsection{Sachs optical equations for a bundle of light rays}
A bundle of light rays is described in General Relativity by the Sachs optical equations \cite{Sachs:1961zz}:
\begin{eqnarray}
\label{Sachs1}
\frac{d\theta}{d\lambda}+\theta^{2}+\bar{\sigma}\sigma= S_R \equiv -\frac{1}{2}R_{ab}k^{a}k^{b}\\
\label{Sachs2}
\frac{d\sigma}{d\lambda}+2\sigma\theta= S_C \equiv C_{abcd}\bar{m}^{a}k^{b}\bar{m}^{c}k^{d},
\end{eqnarray}
where $\lambda$ is an affine parameter along the light ray, $\theta$ and $\sigma$ are, respectively, the isotropic expansion and complex shear scalars of the ray bundle. We have set the vorticity $\omega=0$ because this is always a solution of the equations for a source that radiates isotropically. Alternatively the Sachs equations can be written for the angular diameter distance $r_{A}$ given by:
\begin{equation}
\label{thetara}
\theta=\frac{d\ln r_{A}}{d\lambda}.
\end{equation}
Then they read:
\begin{eqnarray}
\label{Sachsm1}
\frac{1}{r_{A}}\frac{d^{2}r_{A}}{d\lambda^{2}}+\bar{\sigma}\sigma=S_{R}(\lambda)\\
\label{Sachsm2}
\frac{d\sigma}{d\lambda}+\frac{2\sigma}{r_{A}}\frac{d r_{A}}{d\lambda}=S_C(\lambda).
\end{eqnarray} 
The vector field $m^{a}$ in Eq.~(\ref{Sachs2}) is a complex null vector field (we denote by $\bar{m}^{a}$ its complex conjugate) such that $m^{a}m_{a}=0$, $\bar{m}^{a}m_{a}=1$ and $m^{a}k_{a}=0$. It can be decomposed into its real and imaginary parts by introducing two spacelike unit vectors $n_{1}^{a}$ and $n_{2}^{a}$, spanning the screen space orthogonal to the instantaneous direction of propagation of the light rays and parallely transported along the ray: $m^{a}=(n_{1}^{a}-in_{2}^{a})/\sqrt{2}$.
For the basis in the screen space, we can choose, in $(t,x,y,z)$ coordinates, as long as $v_{z}\neq 1$\footnote{Similar expressions can be used if $v_{z}=1$. The system being symmetric by exchange of the axes of symmetry of the lattice (aligned with the coordinate axes), it is enough to consider one particular case; the other ones can be straightforwardly deduced from it.}:
\begin{eqnarray}
\label{VecSP1}
n_{1}^{a}&=&\left(0,\frac{v_x v_z}{\sqrt{1-v_{z}^{2}}},\frac{v_y v_z}{\sqrt{1-v_{z}^{2}}},-\sqrt{1-v_{z}^{2}}\right)\\
\label{VecSP2}
n_{2}^{a}&=&\left(0,-\frac{v_y}{\sqrt{1-v_{z}^{2}}},\frac{v_x}{\sqrt{1-v_{z}^{2}}}, 0\right),
\end{eqnarray}
which have been chosen such that the spatial part of $n_1$ and $n_2$ form an orthonormal basis together with $\mathbf{v}=(v_x, v_y, v_z)$: $\mathbf{n_1} \wedge \mathbf{n_2}= \mathbf{v}$. A calculation using the metric Eq.~(\ref{mainresultmetric}) then yields explicit expressions for the Ricci and Weyl type sources $S_R$ and $S_C$:
\begin{eqnarray}
\label{SR}
S_{R}(\lambda)=-\frac{1}{2}R_{ab}k^{a}k^{b}=\frac{M}{L}\left( -\frac{4 \pi}{L^2}+ \Delta f_{\eta}(\x(\lambda))\right)+\mathcal{O}\left(\frac{M^{3/2}}{L^{3/2}}\right)\\
\label{SC}
S_{C}(\lambda)=C_{abcd}\bar{m}^{a}k^{b}\bar{m}^{c}k^{d}=-\frac{M}{L}\,\frac{\left[i u_z -w_z\right]^2 f_{\eta}(\x(\lambda))}{1-v_z^2} +\mathcal{O}\left(\frac{M^{3/2}}{L^{3/2}}\right),
\end{eqnarray}
where $\mathbf{u}=\mathbf{v} \wedge \nabla$ and $\mathbf{w}=\mathbf{u} \wedge \mathbf{v}$, and where we used to this order $t=\lambda$ and $x^i = v^i \lambda$. We note that the sources for the expansion and the shear, $S_{R}$ and $S_C$, are of order $M/L$. As a consequence, $\bar{\sigma}\sigma$ in Eq.~(\ref{Sachsm1}) is formally of order $M^2/L^2$, so that the shear and the expansion do not couple to each other up to the degree of approximation considered here.  In order to see this more explicitly, let us look for a perturbative solution to Eqs.~(\ref{Sachsm1},~\ref{Sachsm2},~\ref{SR},~\ref{SC}), and expand $r_A= r_A^{(0)}+\sqrt{\frac{M}{L}} \, r_A^{(1)}+ \frac{M}{L} r_A^{(2)} +\mathcal{O}\left(\frac{M^{3/2}}{L^{3/2}}\right)$ and $\sigma= \sigma^{(0)}+\sqrt{\frac{M}{L}} \sigma^{(1)}+ \frac{M}{L} \sigma^{(2)} +\mathcal{O}\left(\frac{M^{3/2}}{L^{3/2}}\right)$. Then the zeroth order obeys Sachs equations in an empty Universe:
\begin{eqnarray}
\label{Sachsvac1}
\frac{1}{r_A^{(0)}}\frac{d^{2}r_{A}^{(0)}}{d^2 \lambda}+\bar{\sigma}^{(0)}\sigma^{(0)}=0\\
\label{Sachsvac2}
\frac{d\sigma^{(0)}}{d\lambda}+\frac{2\sigma^{(0)}}{r_{A}^{(0)}}\frac{d r_{A}^{(0)}}{d\lambda}=0,
\end{eqnarray} 
whose solution is $\sigma^{(0)}=0$ and $r_A^{(0)}= -\lambda$, as in Minkowski spacetime, for the appropriate integration constants, and for sources that radiate isotropically (for which the initial shear is set to zero). First order equations are then found to be:
\begin{eqnarray}
\label{Sachsvaco1}
\frac{d^{2}r_{A}^{(1)}}{d^2 \lambda}=0\\
\label{Sachsvaco2}
\frac{d\sigma^{(1)}}{d\lambda}+2\frac{\sigma^{(1)}}{\lambda}=0.
\end{eqnarray} 
The first order equation for $r_A$ gives $r_{A}^{(1)}=A_1 \lambda +B_1$ and thus only renormalises $r_{A}^{(0)}$. In the following, we will therefore write $r_{A}(\lambda)= -\lambda + \frac{M}{L} r_A^{(2)}(\lambda)$. The second equation is also solved by $\sigma^{(1)}=0$, provided the sources are considered to radiate isotropically. Finally, the second order equations read:
\begin{eqnarray}
\label{Sachsordre21}
\frac{1}{\lambda}\frac{d^{2}r_{A}^{(2)}}{d^2 \lambda}=\frac{4 \pi}{L^2}- \Delta f_{\eta}(\x(\lambda))\\
\label{Sachsordre22}
\frac{d\sigma^{(2)}}{d\lambda}+2\frac{\sigma^{(2)}}{\lambda}=-\frac{\left[i u_z -w_z\right]^2 f_{\eta}(\x(\lambda))}{1-v_z^2}.
\end{eqnarray} 
Hence, as announced, the shear and the expansion (or the angular distance) indeed decouple. More generally, from the form of Sachs equations and the fact that their source is of order $M/L$, it is clear that the shear and the expansion (or angular distance) would couple only at the fourth order in $\sqrt{M/L}$. An analysis of this coupling would require the knowledge of the source terms, and thus of the metric of the lattice Universe, up to $M^2/L^2$ included. This solution is unfortunately out of reach, however, because of the non-linearity of Einstein field equations, that translates in our case into a non-trivial mixing of Fourier modes; see \cite{Bruneton:2012cg}. \\
\\
The fact that the shear and the expansion decouple means that for a lattice Universe where $M/L$ is small, the Weyl curvature does not have \textit{a priori} any noticeable effect on the distance-redshift relation. This is highly unexpected, as one would have thought that, the lattice Universe being mainly empty, the propagation of light would be mostly influenced by the Weyl curvature. In fact, the decoupling is a formal consequence of the expansion in powers of $\sqrt{M/L}$, and thus holds only if this expansion remains a valid approximation. We will show in Section~4 that when the lattice Universe gets too inhomogeneous (i.e. when, for a fixed $M/L$, we consider a more and more peaked distribution for the masses, that is, smaller and smaller $\eta/L$), the expansion in powers of $\sqrt{M/L}$ actually breaks down, in the sense that second order quantities in $\sqrt{M/L}$ happen to be of order $\mathcal{O}(1)$. In this case it is not true any more that shear and angular distance decouple, and the Weyl focussing gets back into the game, as it is expected in a highly inhomogeneous Universe with large and almost empty regions.\\
%
\\
Before going into this discussion, however, we first derive the analytical expression for the angular distance. Eq.~(\ref{Sachsordre21}) is solved by:
$$
r_A^{(2)}(\lambda)= \int_{0}^{\lambda} d \lambda' \int_{0}^{\lambda'} \lambda''\left(\frac{4 \pi }{L^2}- \,\Delta f_{\eta}(\x(\lambda''))\right) d\lambda''.
$$
The integration is best done using the expression for $f_{\eta}$ found in the first line of Eq.~(\ref{feta}). This yields
\begin{eqnarray}
\label{ral1}
r_A(\lambda) &=& -\lambda + \frac{2 \pi}{3}\frac{GM}{Lc^2}\frac{\lambda^3}{L^2} \nonumber \\
&+&\frac{GM}{Lc^2} \sum_{\mathbf{n}\in\Z^{3}_{*}} \left(\frac{i L + \pi \lambda \mathbf{n}.\mathbf{v}}{\pi^2 (\mathbf{n}.\mathbf{v})^3}\right)  e^{-\frac{\pi^{2}|\mathbf{n}|^{2}\eta^{2}}{L^{2}}} e^{ \frac{2 i \pi \lambda \mathbf{n}.\mathbf{v}}{L}},
\end{eqnarray}
 where  $\mathbf{v}=(v_{x}, v_{y}, v_{z})$. This expression only holds if $\mathbf{n}.\mathbf{v} \neq 0$. Let us denote  $\mathcal{S}_{\mathbf{v}}$  the set of triplets $(n,p,q) \in \Z^{3}_{*}$ such that $\mathbf{n}.\mathbf{v} = 0$. The full result is then:
\begin{eqnarray}
\label{ral2}
r_A(\lambda) &=& -\lambda + \frac{2 \pi}{3}\frac{GM}{Lc^2}\frac{\lambda^3}{L^2} 
\nonumber \\
&+&\frac{GM}{Lc^2} \sum_{\mathbf{n}\in \Z^{3}_{*}\setminus \mathcal{S}_{\mathbf{v}}} \left(\frac{i L + \pi \lambda \mathbf{n}.\mathbf{v}}{\pi^2 (\mathbf{n}.\mathbf{v})^3}\right)  e^{-\frac{\pi^{2}|\mathbf{n}|^{2}\eta^{2}}{L^{2}}} e^{ \frac{2 i \pi \lambda \mathbf{n}.\mathbf{v}}{L}} \nonumber \\
&+&\frac{GM}{Lc^2} \sum_{\mathbf{n}\in\mathcal{S}_{\mathbf{v}}}\frac{2 \pi \lambda^3}{3 L^2} e^{-\frac{\pi^{2}|\mathbf{n}|^{2}\eta^{2}}{L^{2}}}.
\end{eqnarray}
 This expression is real valued, and can be further simplified to get, in the end:
\begin{eqnarray}
\label{ral}
r_A(\lambda) &=& -\lambda + \frac{2 \pi}{3}\frac{GM}{Lc^2}\frac{\lambda^3}{L^2} 
\left[1 +\sum_{(n,p,q)\in\mathcal{D}_{\mathbf{v}}} e^{-\frac{\pi^{2}(n^2+p^2+q^2)\eta^{2}}{L^{2}}}\right]\nonumber \\
&+&\frac{2}{\pi }\frac{GM}{Lc^2} \sum_{\mathbf{n}\in \N_{*}^3 \setminus \mathcal{D}_{\mathbf{v}}}  e^{-\frac{\pi^{2} (n^2+p^2+q^2)\eta^{2}}{L^{2}}} \times \nonumber \\
&&\sum_{l=1}^{l=4}  \left[ -\lambda  \frac{\cos\left(\frac{2 \pi \lambda \mathbf{v}.\mathbf{u}_l }{L}\right)}{(\mathbf{v}.\mathbf{u}_l)^2}  +\frac{L}{\pi} \frac{\sin\left(\frac{2 \pi \lambda \mathbf{v}.\mathbf{u}_l }{L}\right)}{(\mathbf{v}.\mathbf{u}_l)^3} \right] \nonumber \\
&+&\mathcal{O}\left(\frac{M^{3/2}}{L^{3/2}}\right).
\end{eqnarray}
Here the $\mathbf{u}_l$'s are the following triplets:
$$
\mathbf{u}_1=(n,p,q), \, \, \mathbf{u}_2=(n,-p,-q), \, \, \mathbf{u}_3=(n,p,-q),\, \, \mathbf{u}_4=(n,-p,q),
$$ 
and $\mathcal{D}_{\mathbf{v}}=\{(n,p,q) \in\N_*^3 : \exists\, l\in\{1,2,3,4\} / \mathbf{u}_l.\mathbf{v}=0\}$. This means that the first sum is over all the triplets that cancel one at least of the $\mathbf{u}_{l}.\mathbf{v}$, whereas the second sum is over all the other triplets.
We note that the first two terms in the angular distance in the first line of Eq.~(\ref{ral}) coincide with their FLRW counterparts for a flat, dust-filled Universe without any cosmological constant, up to order $\lambda^3$, or equivalently up to order $H_0^2 \propto G M/L^3 c^2$. The other terms however shall be characterized more precisely in Section 4.2.

\section{Observables: properties, numerical results and discussion}
So far we derived analytical expressions for the redshift and the angular distance in the lattice Universe, Eqs.~(\ref {redshift}) and (\ref{ral}). Now, we would like to study in more details the 'small' differences between these observables in the lattice, and the same observables in the FLRW model with the same averaged energy density. That is, we would like to quantify the amplitude of the weak backreaction, following the terminology used in  \cite{Kolb:2009rp}, i. e. the difference between the PBS and the GBS, as discussed in the introduction.
\subsection{Redshift}
\begin{figure}[ht]
\begin{center}
\includegraphics[width=10cm]{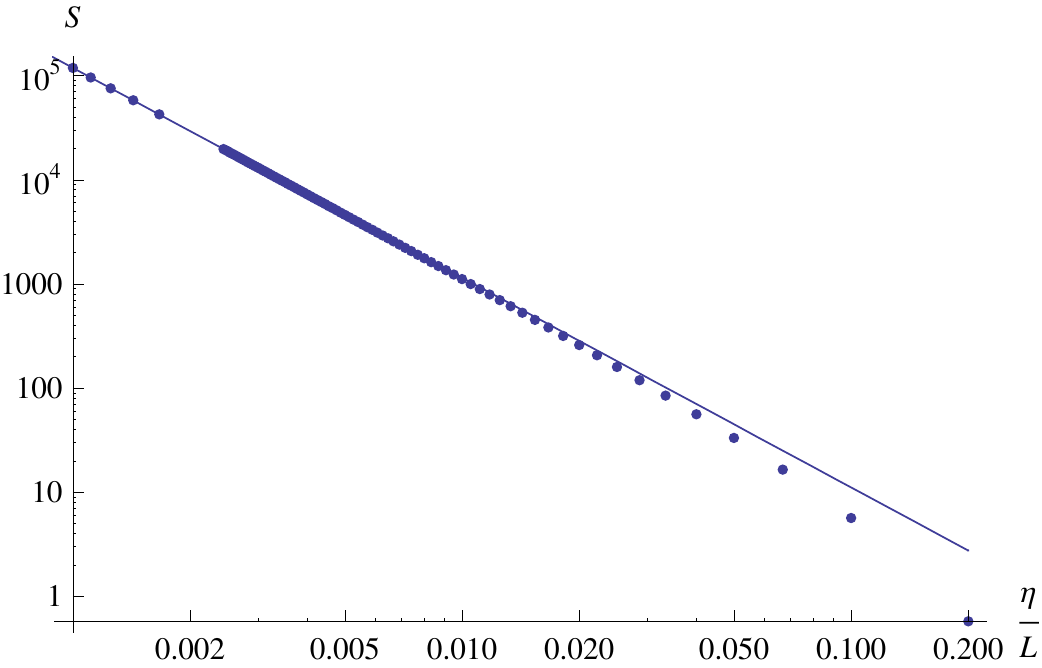}
\end{center}
\caption{Numerical result: Log-Log plot of the sum found in Eq.~(\ref{sommesup}),\\$S=\sum_{(n,p,q)\in\N^{3}_{*}}\left((n +p +q) e^{-\frac{\pi^{2} (n^2+p^2+q^2)\eta^{2}}{L^{2}}}\right)/(n^{2}+p^{2}+q^{2})$, as a function of $\eta/L$, and its best fit of the form $a x^b$ (continuous curve; $a\approx 0.1$, $b \approx -2$). More complicated function would be necessary to capture also the behavior at large $\eta/L$, but the present fit is precise enough in the range of interest for $\eta/L$ (see text).}
\label{thirdterm}
\end{figure}

\begin{figure}[ht]
\begin{center}
\includegraphics[width=0.6\textwidth]{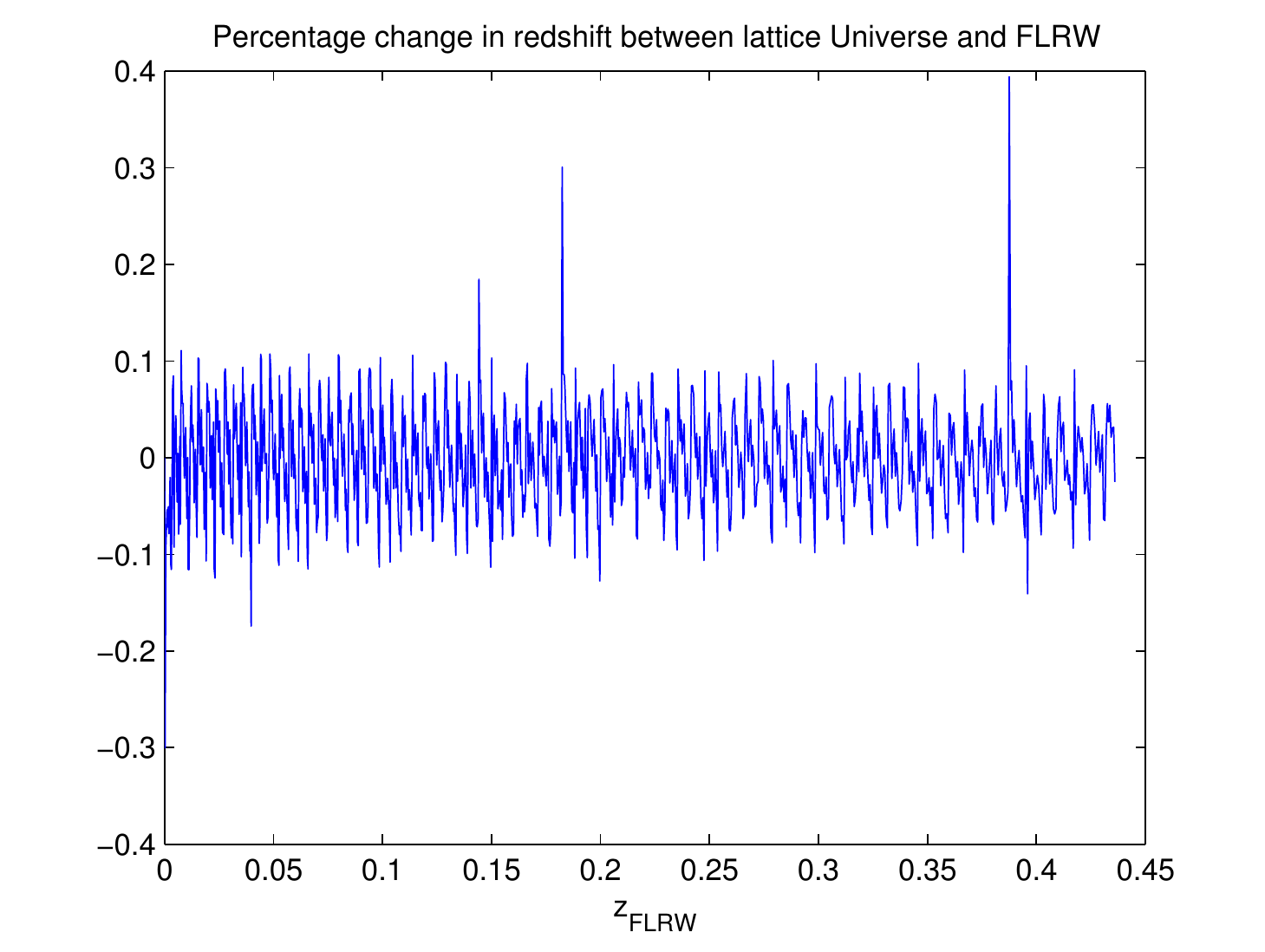}
\end{center}
\caption{Numerical result: Plot of $\delta z/z \equiv (z^{Lattice}-z^{FLRW})/z^{FLRW}$ as a function of $z^{FLRW}$, in percents. These quantities are defined by Eqs. (\ref{redflrw}, \ref{redani}). Here $M/L=10^{-8}$, $\eta/L=10^{-2}$, and only the first $200^3$ terms of the sum are considered.}
\label{deltazsurz}
\end{figure}

The expression for the redshift Eq.~(\ref {redshift}) contains Friedmannian terms coming from the zero mode of the source, given by:
\beq
\label{redflrw}
z^{FLRW}(\lambda)= -\sqrt{\frac{GM}{Lc^2}}\mafrac \frac{\lambda}{L} +\frac{GM}{Lc^2}\frac{14 \pi \lambda^2 }{3 L^2},
\eeq
and anisotropic terms given by:
\beq
\label{redani}
z^{anisotropic}(\lambda)=\frac{GM}{Lc^2}\times\left[ f_{\eta}(\x(\lambda))-\lambda \partial_{i} f_{\eta}(\x(\lambda)) v^i\right]^{\lambda}_{0},
\eeq
coming from the contribution of the other modes. We note that these expressions are only valid for a limited range of $\lambda$ such that $\frac{GM}{Lc^2}\frac{14 \pi \lambda^2 }{3 L^2} \ll \sqrt{\frac{GM}{Lc^2}}\mafrac \frac{\lambda}{L}$. Numerically, this gives
\beq
\label{limitdl}
\frac{|\lambda|}{L} \ll 0.2 \sqrt{\frac{L}{M}},
\eeq
or also, in terms of redshift (putting back into Eq.~(\ref{redflrw})), $z \ll 0.6$. This is the natural limitation of our perturbative approach.\\
\\
Naively, the effect of anisotropies on the redshift must be small for small value of $|\lambda|$ (i.e. small redshifts), since in this case the Friedmannian term in $\sqrt{M/L}$ dominates over the smaller terms or order $M/L$, provided that $M/L \ll 1$. At larger redshifts (or larger $|\lambda|$), the anisotropies must be again subdominant since they behave at most as $M/L \times \lambda$ (the function $f_{\eta}$ being bounded), while the Friedmannian-like term goes as $M/L \times \lambda^2$. However, these conclusions might be affected in some appropriate intermediate range for $\lambda$, if the absolute value of $f_{\eta}$ and $\partial_i f_{\eta} v^i$ is large. Although this is a non-trivial question due to the complicated form of $f_{\eta}$ and its derivative, it is already clear from the expression of $f_{\eta}$ that its magnitude depends on $\eta/L$, i.e. on the width of the sources $\eta$ compared with the separation $L$ between the sources. In other words, the compactness of the lattice plays a major role in fixing the amplitude of the corrections to the pure FLRW behaviour. It is therefore of importance to evaluate how $f_{\eta}$ and its derivative scale with $\eta/L$. \\
\\The simplest term in Eq.~(\ref{redani}) reads
\beq
f_{\eta}(0,0,0)=\frac{8}{\pi}\sum_{(n,p,q)\in\N^{3}_{*}}\frac{e^{-\frac{\pi^{2} (n^2+p^2+q^2)\eta^{2}}{L^{2}}}}{n^{2}+p^{2}+q^{2}}.
\eeq
As a matter of fact, it is difficult to track analytically its magnitude as a function of $\eta$. Numerically, however, we have found the empirical law: $f_{\eta}(0,0,0) \sim 0.73 (\eta/L)^{-1.05}$. Hence, we see that this correction stays small if $ 0.73 (\eta/L)^{-1.05} \ll \frac{14 \pi \lambda^2 }{3 L^2}$. Rounding the figures, and using Eq.~(\ref{limitdl}), this condition reads $\eta \gg 1.2 M$, which always holds since the masses need to be at least as large as their own Schwarzschild radius. There are two other terms in   $z^{anisotropic}$. The first reads $f_{\eta}(\x(\lambda))$, and basically behaves as $f_{\eta}(0,0,0)$ in order of magnitude, since we can consider that the product of the three cosines involved are on average of order unity. The last term reads $\lambda \partial_i f_{\eta} v^i$, and thus behaves as (considering again $\cos \sim 1$ and $\sin \sim 1$)
\beq
\lambda \partial_i f_{\eta} v^i \sim 16 \frac{\lambda}{L}\times \mathcal{O}\left(1\right) \times \sum_{(n,p,q)\in\N^{3}_{*}}\frac{(n v_x+p v_y+q v_z) e^{-\frac{\pi^{2} (n^2+p^2+q^2)\eta^{2}}{L^{2}}}}{n^{2}+p^{2}+q^{2}}.
\eeq
For a given arrival direction of the photon, the $|v^i|$'s are random but of order unity or smaller. Thus we shall bound $\lambda \partial_i f_{\eta} v^i$ by the following sum
\beq
\label{sommesup}
\lambda \partial_i f_{\eta} v^i  \lesssim 16 \frac{\lambda}{L} \times \sum_{(n,p,q)\in\N^{3}_{*}}\frac{(n+p+q) e^{-\frac{\pi^{2} (n^2+p^2+q^2)\eta^{2}}{L^{2}}}}{n^{2}+p^{2}+q^{2}}.
\eeq
Numerically, this sum behaves as $\sim 0.1 (L/\eta)^2$; see Fig.~\ref{thirdterm}. 
%
As a consequence, the corrections due to inhomogeneities remain small with respect to the $M/L$ Friedmannian term provided that 
\beq
\label{bound1}
\frac{M}{L} \ll 0.3 \left(\frac{\eta}{L}\right)^4,
\eeq
all calculations done, and using Eq.~(\ref{limitdl}). For the parameters we chose, namely $M/L = 10^{-8}$ and $\eta/L= 10^{-2}$, this condition is not fulfilled and we should expect large deviations to the FLRW redshift. In fact it is not so (see Fig. \ref{deltazsurz}), because the bound we derived here is only indicative, in the sense that $\lambda \partial_i f_{\eta} v^i$ is typically much less than its superior bound given in Eq.~(\ref{sommesup}). Still, this indicative result roughly determines a subspace in the parameters for which deviations to FLRW cosmology are large\footnote{On evaluating numerically the exact expression for the redshift, we have found that $z^{anisotropic} \sim z^{FLRW}$ when $M/L \sim 1000\, (\eta/L)^4$. This corresponds to the condition for the whole anisotropic contribution to be of order the FLRW one.}. \\
\\
 Interestingly however, one might also show, using the above results, that the corrections due to inhomogeneities (of order $M/L$) remain small with respect to the leading order effect in $\sqrt{M/L}$ if 
\beq
\label{bound2}
\frac{M}{L} \ll 3 \left(\frac{\eta}{L}\right)^4,
\eeq
which is quite close to the previous bound. This shows that large corrections (in the redshift) to the FLRW behaviour are typically associated with a breakdown of our perturbative approach, since then order $M/L$ quantities become comparable to $\sqrt{M/L}$ ones. A very similar conclusion will also arise from the analysis of the behaviour of the angular distance as a function of $\eta$; see next subsection. \\
\\
These bounds Eqs. (\ref{bound1}, \ref{bound2}) (and see also Eq.~(\ref{bound3}) for the angular distance) between the three length scales of the problem, namely $M, \eta$ and $L$, mean the following.  If these conditions are met, meaning essentially that the masses are not too peaked, then the corrections to the observables\footnote{At least for the redshift and the distances.} with respect to the FLRW observables remain small, and this shall be taken, in this case, as an argument in favour of the use of the cosmological principle in standard cosmology (bearing in mind, of course, that we only discuss an approximate solution, with a high degree of symmetry). In other words, the lattice Universe in this case show no strong (dynamical) backreaction, and only small weak backreaction, at least up to the order considered. When these conditions are not met, however, it does not mean \textit{per se} that the FLRW solution is not valid. It only says that the weak backreaction becomes too large: the corrections to the fitting FLRW Universe are so large that the very foundation of our approach to the lattice Universe, namely a series expansion in powers of $\sqrt{M/L}$, cannot be trusted any more. Still, it gives a quantitative estimate of when non-linearities must be taken into account in an inhomogeneous Universe in order to get sensible results. Moreover it shows that the compactness of the sources is a critical parameter impacting the behaviour of the observables in inhomogeneous cosmology, regardless of how small the lattice parameter $M/L$ might be. One must insist, again, on the fact that the lattice is kinematically identical, on average, to the FLRW model with the same energy density. In that sense, there is no strong backreaction 'a la Buchert' \cite{Buchert:1999er,Buchert:2001sa}, irrespective of the compactness of the objects. Nevertheless, our results tend to indicate that there might be weak backreaction (i.e. 'lightcone backreaction'): the fitting FLRW reconstructed from observables might be different from the one reconstructed by simply averaging out spatial inhomogeneities. Only a non-perturbative treatment, solving exactly the coupled Sachs equations, could answer this question.\\
\\
All the previous analysis for the redshift has been checked against direct numerical calculation of $z^{anisotropic}$, displayed below. In Fig.~\ref{deltazsurz} we plot $(z^{Lattice}-z^{FLRW})/z^{FLRW} = z^{anisotropic}/z^{FLRW}$ as a function of the FLRW redshift for $M/L=10^{-8}$, $\eta/L=0.01$,  and for a randomly chosen arrival direction $\mathbf{v}$. The corrections are less than one percent, although we have checked that the discrepancy with respect to the homogeneous Universe increases with the degree of inhomogeneity (i.e. for decreasing $\eta/L$), as expected in the light of the previous discussion. The amplitude of the effect does not depend significantly on the $v^{i}$'s selected. In order to perform numerical calculations, we have truncated the sum defining the function $f_{\eta}$ or its derivative, and summed over non-null triplets $(n,p,q) \in \{0,...,N\}^3$ with $N=200$, which is enough to ensure convergence with such a value of $\eta/L$. The noisy character comes from the fact that the photon reaching the observer has to escape the local gravitational well from where it is emitted. Notice that such a noise induces a natural spreading of the luminosity distance $d_L(z)$ around its main Friedmannian value, which is not due to the peculiar velocities of the sources, but is rather an intrinsic effect coming from the existence of local gravitational wells, and which moreover couples to the expansion, i.e. it is a kind of Sachs-Wolfe effect. For this reason, such a spreading of the Hubble-Sandage diagram shall actually be shared by any inhomogeneous models. The amplitude and the amount of spread for the redshift can be evaluated in our case, although it stays below observable limits for realistic values of the parameters. 
 
\subsection{Angular distance}
\begin{figure}[ht]
\begin{center}
\includegraphics[width=\textwidth]{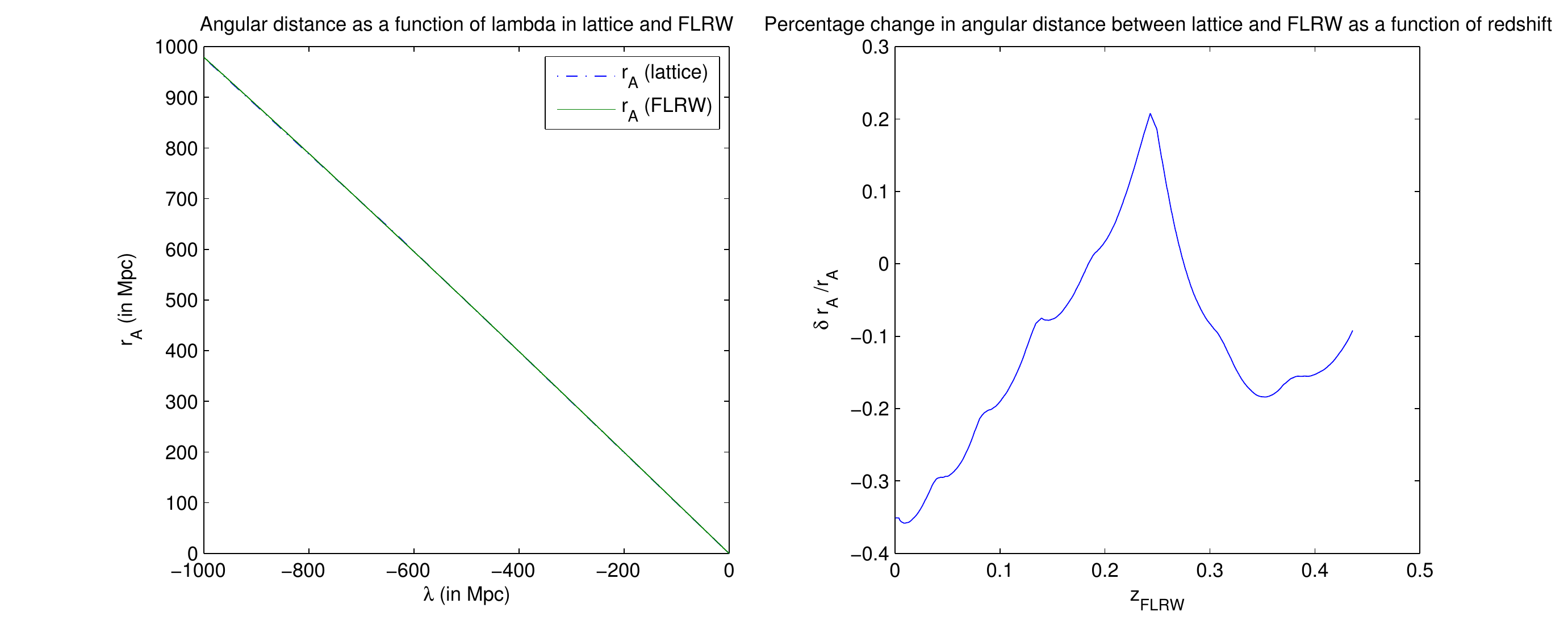}
\end{center}
\caption{Left: The angular distance as a function of the affine parameter $\lambda <0$ in the lattice Universe (dashed line) and FLRW (continuous curve), for $M/L=10^{-8}$ and $\eta/L=0.03$, i.e.  $M/L \approx 0.012 (\eta/L)^4$, and random $v^{i}$'s. The order of magnitude of the effect does not depend on the $v^{i}$'s selected. The plot is based on $1500$ calculated points, and the cutoff for the sums is $N=200$. The two lines are indistinguishable. Indeed, the relative difference is less than $0.4 \%$; see right panel.}
\label{ra03}
\end{figure}

\begin{figure}[ht]
\begin{center}
\includegraphics[width=\textwidth]{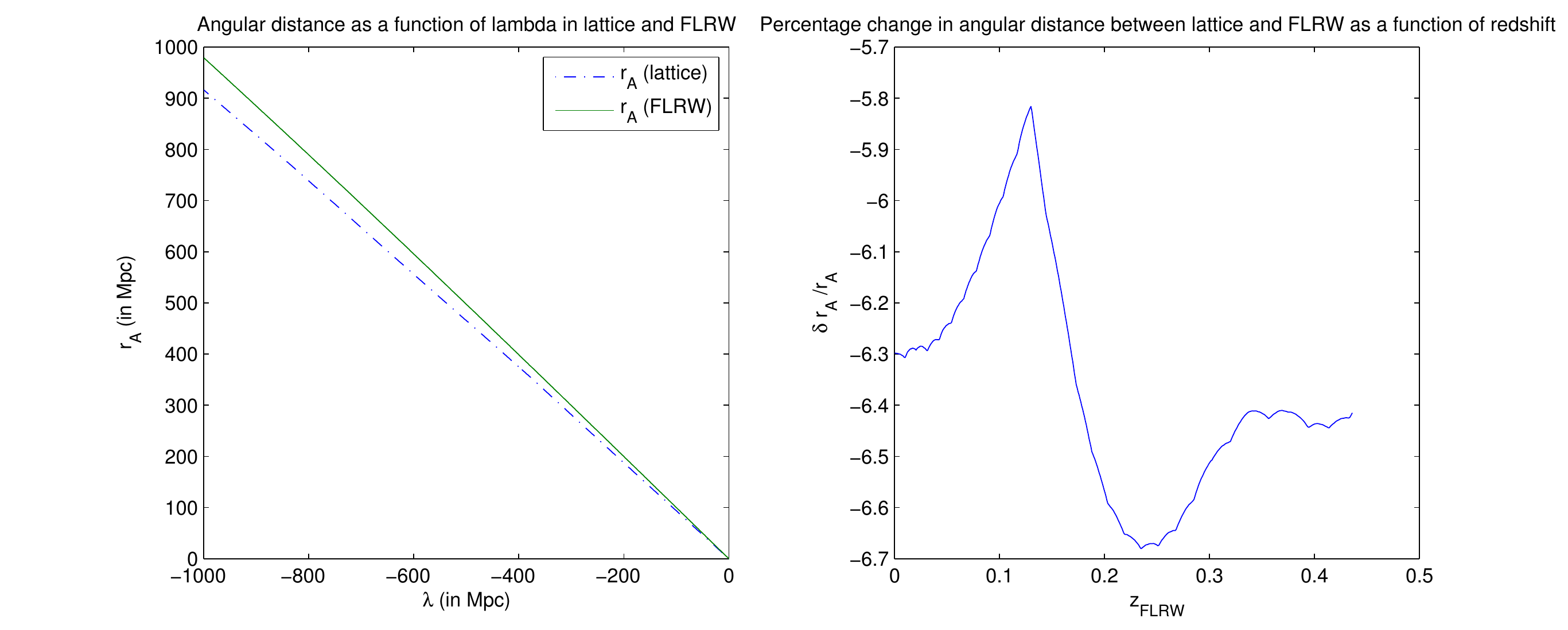}
\end{center}
\caption{Left: The angular distance as a function of the affine parameter $\lambda <0$ in the lattice Universe (dashed line) and FLRW (continuous curve), for $M/L=10^{-8}$ and $\eta/L=0.017$, i.e.  $M/L \approx 0.12 (\eta/L)^4$, and random $v^{i}$'s. The order of magnitude of the effect does not depend on the $v^{i}$'s selected. The plot is based on $500$ calculated points, and the cutoff for the sums is $N=200$. The angular distance in the lattice Universe is significantly less than the FLRW distance, due to the negative contribution of the 'quasi-poles', as discussed in the text. The relative difference is around $6 \%$; see right panel. This illustrates the sharp transition in the behaviour of the observables as a function of $M/L\times (L/\eta)^4$.}
\label{ra017}
\end{figure}

The equation for the angular distance, Eq.~(\ref{ral}), shows two types of corrections to the FLRW behaviour, depending on whether the triplets $(n,p,q) \in\N_*^3$ involved are poles (such that $\mathbf{u}_l.\mathbf{v}=0$ for some $l$) or not. Let us first focus on the first correction to FLRW, reading: 
\beq
r_A(\lambda)=-\lambda + \frac{2 \pi}{3}\frac{GM}{Lc^2}\frac{\lambda^3}{L^2} 
\left[1 +\sum_{(n,p,q)\in\mathcal{D}_{\mathbf{v}}} e^{-\frac{\pi^{2}(n^2+p^2+q^2)\eta^{2}}{L^{2}}}\right]+ \ldots,
\eeq
where we recall that $\mathcal{D}_{\mathbf{v}}=\{(n,p,q) \in\N_*^3 : \exists\, l\in\{1,2,3,4\} / \mathbf{u}_l.\mathbf{v}=0\}$, see Eq.~(\ref{ral}) and below. This set is given by the intersection of the plane whose normal is $\mathbf{v}$, with the comoving lattice of masses. The sum over $\mathcal{D}_{\mathbf{v}}$ is thus very difficult to perform in general. However, we can expect the sum to be negligible for general (random) values of the $v^i$'s, since then we expect $\mathcal{D}_{\mathbf{v}}$ to be sparsely distributed in $\N_*^3$. The sum involves many terms only for some very specific values of $\mathbf{v}$ corresponding to the symmetries of the cube. Hence we can evaluate that the sum is always less than its particular value for a photon propagating along the axes of the lattice, for example, $v_x=v_y=0$, $v_z=1$, in which case $\mathcal{D}_{\mathbf{v}}=\{(n,p,0): (n,p) \in \N_*^2\}$. Then the sum can be computed analytically in terms of the Jacobi elliptic function:
\beq
1+\sum_{(n,p)\in\N_*^2} e^{-\frac{\pi^{2}(n^2+p^2)\eta^{2}}{L^{2}}}=\frac{1}{4}\left(1+\nu\left(e^{-\eta^2/L^2}\right) \right)^2,
\eeq
where $\nu(t)=1+2 \sum_{n=1}^{n=\infty} t^{n^2}$. Numerically, this behaves as $\sim \mathcal{O}(1)\times (\eta/L)^{-2}$. This means that, at most, and in very rare cases, the corrections to the $\lambda^3$ term can get quite large, in which case this significantly reduces the range of validity of the perturbative scheme followed here:
\beq
|\lambda| \gg \frac{2 \pi}{3}\frac{GM}{Lc^2} \frac{|\lambda|^3}{L^2}\times \mathcal{O}\left(\frac{L}{\eta}\right)^2 \Leftrightarrow |\lambda| \ll \mathcal{O}\left(\frac{\eta}{L}\right) \sqrt{\frac{L}{M}},
\eeq
instead of the natural range $\frac{|\lambda|}{L} \ll 0.2 \sqrt{\frac{L}{M}}$ found in Eq.~(\ref{limitdl}). For an  inhomogeneous Universe with $\eta=10^{-2} L$, this reduces the range of attainable redshifts to already uninteresting values of order $10^{-2}$, thus illustrating again the inadequacy of the perturbative expansion when evaluating observables in the lattice Universe. However, this calculation only holds for very specific values of $\mathbf{v}$. We do not expect such large corrections in general. Still, this correction reduces in any case the range of applicability of the perturbation scheme followed here.\\
\\
The second correction to FLRW in Eq.~(\ref{ral}) is more interesting. This expression involves pole-like terms proportional to $1/(\mathbf{u}_l.\mathbf{v})^2$.
%
 Although $\mathbf{u}_l.\mathbf{v}$ cannot vanish strictly by definition of $\mathcal{D}_{\mathbf{v}}$, it might still get very large in general. Hence we shall talk about 'quasi-pole' in the following. When such a 'quasi-pole' $1/(\mathbf{u}_l.\mathbf{v})^2 \gg 1$ occurs, it basically dominates over all the other terms in the sum in Eq.~(\ref{ral}), in which case the angular distance simplifies to:
\begin{eqnarray}
\label{ralpole}
r_A(\lambda)&\sim&-\lambda + \frac{2 \pi}{3}\frac{GM}{Lc^2}\frac{\lambda^3}{L^2} 
\left[1 +\sum_{(n,p,q)\in\mathcal{D}_{\mathbf{v}}} e^{-\frac{\pi^{2}(n^2+p^2+q^2)\eta^{2}}{L^{2}}}\right]\nonumber \\
&+& \lambda \frac{2}{\pi}\frac{GM}{Lc^2} \frac{e^{-\frac{\pi^{2} (n_{*}^2+p_{*}^2+q_{*}^2)\eta^{2}}{L^{2}}}}{(\mathbf{v}.\mathbf{u}_{l_{*}})^2},
\end{eqnarray}
where only the largest 'quasi-pole' given by some triplet $(n_{*}, p_{*}, q_{*})$ and a specific value for $l_{*}$ is considered here. In deriving the previous expression, we also used that $\cos\left(\frac{2 \pi \lambda \mathbf{v}.\mathbf{u}_{l_{*}} }{L}\right) \sim 1$ and $\sin\left(\frac{2 \pi \lambda \mathbf{v}.\mathbf{u}_{l_{*}} }{L}\right) \sim \frac{2 \pi \lambda \mathbf{v}.\mathbf{u}_{l_{*}} }{L}$ since $\mathbf{v}.\mathbf{u}_{l_{*}} \ll 1$. We note that the angular distance is thus typically smaller than the one in the kinematically fitting FLRW Universe since $\lambda <0$. This was confirmed by direct numerical calculation including all the terms in the sum; see below. It can even happen that the angular distance becomes negative, if the $v^i$'s are such that there exist triplets for which the pole is large enough. This can only happen when the second order term above (in $M/L$) gets as large as the zeroth order term $-\lambda$. Such an unphysical result must therefore not be seen as a catastrophe \textit{per se}, but rather as a signal that our perturbative scheme breaks down a at certain point\footnote{From Eq.~(\ref{thetara}), we have $r_a= \exp \left(\int^\lambda \theta(\lambda) d\lambda\right)$, which must be positive. However, when truncated at order $M/L$, the angular distance needs not be positive anymore. In this case, higher order terms need to be considered, signaling the failure of the perturbative expansion used here.}.\\
\\
In order to get a quantitative estimate of the range of validity of our perturbative expansion, we have thus studied the probability distribution of the magnitude of the 'quasi-pole' terms  $1/(\mathbf{u}_l.\mathbf{v})^2$ as a function of $\eta/L$. Our method was the following. For given random values for the $v^i$'s on the unit sphere, we looked for the maximal value of $1/(\mathbf{u}_l.\mathbf{v})^2$ for all $l=1,\ldots,4$, and all triplets $(n,p,q)$ in $\{0,...,N\}^3$ such that $(n,p,q)\neq (0,0,0)$. We repeated the calculation for a large number ($1000$) of random values for the $v^i$'s, and thus ended up with a large enough distribution of the maximal value of the 'quasi-pole' as a function of $N$. We thus took the median of this set to find the typical value of the largest 'quasi-pole' as a function of $N$. Now varying $N$, we found numerically that the typical size of the largest 'quasi-pole' as a function of $N$ goes like $\sim 22 \,N^{4.1}$. The main correction $\delta r_A/r_A$ to Friedmann's law for the angular distance then reads, using the previous result and Eq.~(\ref{ralpole}):
\beq
\left|\frac{\delta r_A}{r_A}\right| \sim  \frac{44}{\pi}\frac{GM}{Lc^2} N^{4.1} e^{-\frac{\pi^{2}N^2\eta^{2}}{L^{2}}}.
\eeq
In order for the perturbative expansion to remain a  valid approximation, we require $\left|\frac{\delta r_A}{r_A}\right| \ll 1$. Rounding the exponent to $4$, the function $N^{4} e^{-\frac{\pi^{2}N^2\eta^{2}}{L^{2}}}$ is maximal for $N=\sqrt{2}L/(\pi \eta)$ and its maximal value reads $(\eta/L)^{-4}/12.8$. Rounding again the figures, we thus find that our perturbative approach to the calculation of observables in the lattice Universe only holds if:
\beq
\label{bound3}
\frac{M}{L} \ll 10 \left(\frac{\eta}{L}\right)^4,
\eeq
all calculations done. This bound defines a restricted range of validity of our perturbative approach, independent of $\lambda$, besides the natural limitation Eq.~(\ref{limitdl}) of the Taylor expansion in powers of $\sqrt{M/L}$. If the free parameters satisfy Eq.~(\ref{bound3}) above, then it is very likely that the pole-like terms do not blow up. In this case then, the corrections to the FLRW angular distance will be small. On the other hand, if (\ref{bound3}) is not satisfied, it is very likely that a large 'quasi-pole' shows up, in which case the corrections are large and can even lead to unphysical results such as a negative angular distance, meaning indeed a failure of the perturbative expansion.  The relevance of the above bound has been checked against full numerical computation of the angular distance. It turns out numerically that the equation $\frac{M}{L} \sim 10 \left(\frac{\eta}{L}\right)^4$ is overestimated, and that the transition between the two limiting behaviours between small and large corrections to FLRW occurs more precisely around $\frac{M}{L} \sim 0.1 \left(\frac{\eta}{L}\right)^4$; see Figs.~\ref{ra03} and \ref{ra017} for an illustration of the two cases of small and large corrections. \\
\\Interestingly, this roughly corresponds to the bound found for the redshift\footnote{Although, as explained in Section 4.1, the strength of the corrections to $z^{FLRW}$ have been overestimated in the analytical treatment.}, in particular concerning the exponent, $4$. More generally then, we can conclude our analysis of Section 4.1 and 4.2, by saying that the general equation
\beq
\label{mainbound}
\frac{M}{L} \ll \mathcal{O}(1) \times \left(\frac{\eta}{L}\right)^4,
\eeq
relating the lattice parameter $M/L$ to the extension of the sources $\eta/L$ is the critical condition for the corrections to FLRW to stay small, while, under this condition, the perturbative expansion followed here stays under control. Although we only deal with an approximate solution for an idealized lattice, this result suggests that the effects of inhomogeneities upon observable quantities in more realistic Universes are to stay small provided that the above bound is met. What happens in a more inhomogeneous Universe where Eq.~(\ref{mainbound}) is not satisfied is a more difficult question. We saw that it implies \textit{a priori} large differences between the PBS and the GBS, at second order in a perturbative expansion in terms of $\sqrt{M/L}$. However, in this case, it turns out that the perturbative expansion breaks down, leaving the possibility for higher order terms to compensate for second order corrections, and more generally, to compensate between each others. This goes beyond the scope of this paper, since this study would require a solution to Einstein equations up to higher orders, and a solution to the Sachs equations that couple shear and isotropic expansion.\\
\\
Note that this effect (of the 'quasi-poles') is new and cannot be accounted for by a lensing effect due to the fact that the light ray travels too close to a mass. Indeed, we have seen that the probability of having a large 'quasi-pole' in the sum is inversely proportional to the fourth power of the size of the objects. But, one would expect the probability for a light ray to travel close to a mass to be proportional to the projected area of the objects in the planes orthogonal to the direction of propagation, i.e., to be proportional to the square of the size, and to be more important for less compact object; this is the converse of the effect of the 'quasi-poles'. Therefore, we must conclude that this effect is truly a problem of the perturbative expansion: as we will see in the next subsection, when the bound (\ref{bound3}) is not satisfied, the shear generated by the Weyl curvature is no longer of order $M/L$ and cannot be neglected in the Sachs equation for $r_{A}$ anymore. In a way, this will establish the relation (\ref{bound3}) as a criterion to determine whether or not, in a given Universe, the Weyl curvature plays any role in the propagation of light.


\subsection{Effect of the Weyl curvature; break-down of the perturbative expansion}
The shear is given by $\sigma = \frac{M}{L} \sigma^{(2)}$ where $\sigma^{(2)}$ is given by the differential equation (\ref{Sachsordre22}). The equation is formally solved by:
\beq
\sigma^{(2)}(\lambda)=-\frac{1}{\lambda^2} \int \lambda^2 \frac{\left[i u_z -w_z\right]^2 f_{\eta}(\x(\lambda))}{1-v_z^2}d \lambda
\eeq
Provided $\mathbf{n}.\mathbf{v} \neq 0$, a calculation shows that:
\begin{eqnarray}
\sigma&=& -\frac{i}{\pi^2 L(1-v_z^2)}\frac{GM}{Lc^2} \frac{L^2}{\lambda^2} \sum_{(n,p,q)\in\Z^{3}_{*} / \mathbf{n}.\mathbf{v}\neq 0}\frac{ e^{-\frac{\pi^{2} \mathbf{n}^2\eta^{2}}{L^{2}}} e^{\frac{2 i \pi \lambda \mathbf{n}.\mathbf{v}}{L}}}{\mathbf{n}^2 (\mathbf{n}.\mathbf{v})^3 }\times \nonumber \\
&&\left( -1 +  2 i\pi \frac{\lambda}{L}\mathbf{n}.\mathbf{v} + 2 \pi^2\frac{\lambda^2}{L^2}(\mathbf{n}.\mathbf{v})^2\right)\left( i U_z- W_z\right)^2 \nonumber \\
&+&\mathcal{O}\left(\frac{M^{3/2}}{L^{3/2}}\right).
\label{ShearBlow}
\end{eqnarray}
where $\mathbf{U} \equiv \mathbf{v} \wedge \mathbf{n}$ and $\mathbf{W} \equiv \mathbf{U} \wedge \mathbf{v}$. The case $\mathbf{n}.\mathbf{v}=0$ leads to terms proportional to $\lambda$. This expression is only defined when $v_{z}\neq 1$; if $v_{z}=1$, a similar expression could be found by changing the basis in the screen space ({\it cf} footnote before Eqs.~(\ref{VecSP1}) and (\ref{VecSP2})).
The precise expression does not matter much here. What is important is that we thus find similar terms in the shear and in the angular distance, especially the poles and 'quasi-poles'in $\mathbf{n}.\mathbf{v}$. We saw in the previous section that when such 'quasi-poles' blow up, that is, when they become numerically of order $L/M \gg 1$, which is very likely if the bound $\frac{M}{L} \ll \mathcal{O}(1) \times \left(\frac{\eta}{L}\right)^4$ is not met, then the corrections to the angular distance due to inhomogeneities become very large and the perturbative expansion breaks down. It is thus very interesting to note that the shear itself becomes large around the same time and that it is no longer of order $M/L$. In this case, the term $\bar{\sigma}\sigma$ appearing in the Sachs equation for the angular distance is not of order $M^2/L^2$ anymore, and therefore cannot be neglected. Thus, the bound Eq.~(\ref{mainbound}) is also critical as to assess whether the shear and angular distance decouple or not at first orders. This shows that, as expected, the Weyl curvature plays an important role in very inhomogeneous Universes violating this bound. As a matter of fact the precise point at which the shear becomes large cannot simply be read out of Eq.~(\ref{ShearBlow}), because this expression was obtained by neglecting the contribution of the angular distance to the equation for the shear; but we saw that this angular distance becomes large when the bound Eq.~(\ref{mainbound}) is not satisfied, therefore, a full solution for the shear must take into account an $r_{A}$ of order one. This emphasises the fact that, in order to have a good understanding of the amplitude of the weak backreaction in the lattice when the bound Eq.~(\ref{mainbound}) is not met, one must solve the full system of coupled Sachs equations, without any perturbative scheme, even though the metric can still be written perturbatively. This is beyond the scope of the current work and is left for a future study.

%
\section{Conclusion}
In this paper, we showed that in a lattice Universe kinematically equivalent to an FLRW model with the same averaged energy density, thus showing no 'strong backreaction' in the sense of \cite{Kolb:2009rp}, angular and luminosity distances would not deviate significantly from the ones in the FLRW model provided the spread $\eta/L$ of the object was related to the mass of the objects $M/L$ via:
\beq
\label{mainboundConclusion}
\frac{M}{L} \ll \mathcal{O}(1) \times \left(\frac{\eta}{L}\right)^4.
\eeq
In our model, compactness is thus a key parameter in regards of the amplitude of the 'weak backreaction' \cite{Kolb:2009rp}. In other words, this relation can thus be understood as a criterion to decide whether or not the fitting problem is a problem at all within these models: if this relation is satisfied, observables are almost the ones of the kinematically fitting FLRW model. Otherwise, if objects are too compact, perturbative estimates are not reliable, and it is impossible to say how observables relate to the fitting model without a fully non-perturbative treatment of the propagation of light. This should not be understood as a claim that there is a genuine fitting problem in our 'real' Universe, but simply as a warning that there exist spacetime configurations such that, despite the fact that the solution of Einstein field equations remains close to an FLRW configuration, observables might significantly deviate from their FLRW analogue, or at least, need to be calculated non-perturbatively; see \cite{Clarkson:2011uk} for a similar result in FLRW plus perturbations. Indeed, we hinted at a link between this break-down of the perturbative expansion and the fact that the Weyl curvature behaves non perturbatively when the bound (\ref{mainboundConclusion}) is not satisfied, leading to big corrections to the angular distance at second order in $\sqrt{M/L}$ that might be compensated non perturbatively by the effect of the shear. It would be interesting to check if similar conclusions can be drawn in more 'realistic' configurations of the matter distribution. In particular, provided one could find a satisfactory way of estimating the compactness of cosmological objects, it would be interesting to see whether or not the bound (\ref{mainboundConclusion}) is satisfied in our Universe, using galaxy surveys and/or N-body simulations. A future work must also present the numerical solution of the full system of Sachs equations non perturbatively, in order to avoid the limitations of the bound (\ref{mainboundConclusion}); only such a solution will allow one to decide what happens when this bound is not satisfied. The results of \cite{Holz:1997ic} imply, through statistical arguments, that there is no fitting problem in cosmology, even for a matter distribution made of point masses, except for the exceptional light-rays that travel too close to some masses. Our model, with a high degree of symmetry, seems to indicate otherwise: the 'quasi-poles' seem to act only in one way, decreasing the angular distance compared to its FLRW counter-part, and their effect cannot be accounted for by a simple encounter of the light ray with the neighbourhood of a mass. Nevertheless, it is actually impossible to conclude and one will need the full, non perturbative, solution in order to do some statistics and to tests the results of \cite{Holz:1997ic} with the lattice solution presented here. This is the subject of an ongoing investigation.
 
\ack
J.-P. B. is FSR/COFUND postdoctoral researcher at naXys. J.-P. B. and J. L. acknowledge fruitful discussions with T. Clifton while preparing this paper. We thank anonymous referees for their valuable comments on an earlier version of this paper.

\section*{References}
\bibliographystyle{unsrt}
\bibliography{Torus_biblio}
\end{document}